\documentclass[sigconf]{acmart}
\usepackage{enumitem}
\usepackage{soul}
\usepackage{pifont}
\usepackage{breakurl}
\usepackage[utf8]{inputenc}

\UseRawInputEncoding

\copyrightyear{2022}
\acmYear{2022}
\setcopyright{rightsretained}
\acmConference[CHI '22]{CHI Conference on Human Factors in Computing Systems}{April 29-May 5, 2022}{New Orleans, LA, USA}
\acmBooktitle{CHI Conference on Human Factors in Computing Systems (CHI '22), April 29-May 5, 2022, New Orleans, LA, USA}
\acmDOI{10.1145/3491102.3517546}
\acmISBN{978-1-4503-9157-3/22/04}
\begin{document}

\title[A Little Too Personal: Communication Strategies in Online Freelancing]{A Little Too Personal: Effects of Standardization versus Personalization on Job Acquisition, Work Completion, and Revenue for Online Freelancers}

\author{Jane Hsieh}
\affiliation{%
  \institution{Carnegie Mellon University}
  \city{Pittsburgh}
  \state{Pennsylvania}
  \country{USA}
}
\email{jhsieh2@cs.cmu.edu}

\author{Yili Hong}
\affiliation{%
  \institution{University of Houston}
  \city{Houston}
  \state{Texas}
  \country{USA}}
\email{yilihong@uh.edu}

\author{Gordon Burtch}
\affiliation{%
  \institution{Boston University}
  \city{Boston}
  \state{Massachusetts}
  \country{USA}}
\email{gburtch@bu.edu}

\author{Haiyi Zhu}
\affiliation{%
  \institution{Carnegie Mellon University}
  \city{Pittsburgh}
  \state{Pennsylvania}
  \country{USA}}
\email{haiyiz@cs.cmu.edu}

\renewcommand{\shortauthors}{Hsieh, et al.}

\newcommand{\red}[1]{#1}
\newcommand{\final}[1]{#1}

\begin{abstract}
  As more individuals consider permanently working from home, the online labor market continues to grow as an alternative working environment. While the flexibility and autonomy of these online gigs attracts many workers, success depends critically upon self-management and workers' efficient allocation of scarce resources. To achieve this, freelancers may develop alternative work strategies, employing highly standardized schedules and communication patterns while taking on large work volumes, or engaging in smaller numbers of jobs whilst tailoring their activities to build relationships with individual employers. In this study, we consider this contrast in relation to worker communication patterns. We demonstrate the heterogeneous effects of standardization versus personalization across different stages of a project and examine the relative impact on job acquisition, project completion, and earnings. Our findings can inform the design of platforms and various worker support tools for the gig economy.
\end{abstract}



\maketitle

\section{Introduction}

Online freelancing platforms employ gig workers and service clients around the globe to accomplish virtually deliverable tasks spanning a wide range of job categories and expertise levels. For many workers, the online gig economy provides an alternative to the traditional workplace, one that offers more autonomy, mobility, and flexibility - both physically and temporally \cite{dunn2017digital, boundary}. These affordances attracted 59 million workers in 2020, more than a third of the US workforce. Those workers collectively earned \$1.2 trillion and 36\% of them participated on a full-time basis \cite{freelance-forward}. In 2019, online freelancing allowed 46\% of the gig worker population to be employed despite personal circumstances (e.g. caretaking, disabilities, etc.), and in 2016 it served as a primary source of income for 44\% of workers \cite{manyika2016independent}.

However, the touted benefits of gig work come at a cost. On the other side of the coin, \red{decisional} autonomy implies self-management, an overhead that is otherwise absent in traditional employment\red{, while physical and temporal flexibility may result in a lack of boundaries between work and personal time - a phenomenon known as the autonomy paradox \cite{Shevchuk_Strebkov_Davis_2019}}. Gig workers must efficiently manage their limited resources such as energy, time and, connections.

To achieve lateral mobility (the freedom to work across different career fields or platforms), freelancers must take extra initiative to engage in skill training and invest more hours to seek out jobs across various sectors \cite{boundary}. One particularly challenging aspect of online gig work is the digitally-mediated nature of communication. Although a remote working arrangement can enable temporal and spatial flexibility, digitally-mediated labor also deprives workers of many benefits inherent in face-to-face communication. Messages relayed through online channels suffer from a plethora of complications, including asynchronicity, connectivity issues, time zone differences, and a lack of nonverbal cues (e.g. tone of voice and body language such as gestures and facial expressions). 

To cope with these challenges and efficiently manage their efforts, gig workers may develop alternative strategies, such as \textbf{\textit{standardization}} or \textbf{\textit{personalization}}. Standardization work patterns may involve, for example, the use of job proposal templates to quickly submit multiple bid applications to different projects. It may also involve the use of a fixed working schedule. Personalization, in contrast, may entail customizing the content of a bid proposal to cater toward a particular client's needs, or tailoring a work schedule to align with or accommodate that of an employer.

In this paper, we empirically investigate the efficacy of the above strategies, seeking answers to the following research questions: 1) How does standardization versus personalization in initial employer communications influence a freelancer's likelihood of winning a job? 2) How does standardization versus personalization in the timing of a worker's communication influence a freelancer's likelihood of completing an awarded project? and 3) How do these practices impact workers' broader earning efficiency in the market? Using data from a leading global freelancing platform, we analyze communication patterns derived from 2,031,068 direct messages exchanged between 58,397 freelancers and 25,480 employers, in relation to 56,222 projects, between January and March of 2010. We provide evidence that 1) personalizing initial communications toward a particular job increases the likelihood of being hired 2) maintaining a consistent work schedule increases the likelihood of project completion and 3) \red{content} standardization enables greatest overall earnings in the market, by allowing the freelancer to have larger bid and work volumes.

With the recent investment in and shift toward remote work by various workers and organizations, there is reason to believe that the prevalence of online gig work will continue to rise \cite{Barrero2020-sf, huang2020unemployment}. Our findings are thus important, as they contribute to our understanding of effective work and communication strategies that may enable new freelancers to acclimate and succeed in this \red{novel} labor environment \red{- and distinct from other approaches common in HCI since it takes a more quantitative approach toward human-centered optimization.} We discuss the practical implications that our empirical findings can have on the design of gig platforms \red{as well as} worker support tools that aim to assist freelancers in maximizing their working efficiencies and individual well-beings.

\section{Related Work}

\subsection{\final{Challenges Endemic to Online Gig Work}}

Gig work has been defined as ``electronically mediated employment arrangements in which individuals find short-term tasks or projects via websites or mobile apps that connect them to clients and process payment" \cite{Kuhn2019-fi}. 
\red{In short, online labor platforms act as an intermediary (providing boundary resources such as communication channels, evaluation metrics and automated transactions \cite{platform_manage}) between freelancers who seek jobs and clients who look to hire professionals to complete various forms of work.} 
\red{ In this study we focus primarily on virtually-deliverable and knowledge-intensive work, as opposed to physical services such as furniture assemblage offered by gig workers on platforms such as TaskRabbit, ridesharing services that are now commonly provided by Uber/Lyft, or microwork such as those found on Amazon Mechanical Turk.} 
The digital nature of gig work \red{suggests} the prospect\red{s} of greater work flexibility and independence, and many workers report pursuing gig work with these benefits in mind \cite{Hall2018-uo}. However, \red{such affordances} comes at the cost of several unique challenges. For example, gig workers may face professional isolation and atomization over the long term \red{\cite{Yao_Weden_Emerlyn_Zhu_Kraut_2021}}, \red{causing them to} obtain fewer networking and \red{advancement} opportunities (reduced lateral \red{and upward} mobility\red{, respectively}) \red{ as well as limited social support while enduring fierce competition} \cite{boundary}. 
\red{Beyond social isolation,} workers must also contend with unique day-to-day difficulties, such as income instability and the need to self-manage, e.g. coordinating their time and resources, \red{maintaining productivity,} self-advertising, proactively seeking out new work, \red{building reputation,} and maintaining client relationships \red{\cite{Sutherland2020-wk}}.

\red{The unique structure of online labor platforms introduces elements of uncertainty and information asymmetries (sometimes intentionally), which can create power imbalances that favor clients and enable platformic management \cite{platform_manage, power, power_struggles, Sutherland2020-wk, Rosenblat_Stark_2016, Jarrahi_Sutherland_2019, kingsley2015accounting}. For instance, clients on Upwork are not required to disclose their identities (a privilege not afforded to its freelancers) and they may also leave private reviews for workers they've hired; freelancers, on the other hand, cannot even access the other bidders of a project they apply to, nor can they see who the ultimate winner is \cite{platform_manage}. Specific components of platform structure, such as calculated ratings, have also been found to increase power asymmetries and worker precarity \cite{Sutherland2020-wk}.} 

\red{Compounding on their already} precarious job situation\red{s, freelancers} can be highly susceptible to volatility in the marketplace. For instance, Huang et al. \cite{huang2020unemployment} found that, during the 2008 recession, an unemployment increase of 1\% was associated with a 14.9\% increase in project bidding and a 21.8\% rise in the number of active workers. More recently,  Sutherland et al. \cite{Sutherland2020-wk} discovered that the COVID-19 pandemic caused a decrease in worker-controlled flexibility, along with increased competition, exploitation and workload intensity. 
As we gradually transition from this era of work from home, many employees face the dilemma of whether to remain remote \cite{Barrero2020-sf}. Some companies hesitate to offer such long-term remote work options for their employees, which urges many workers to turn to freelancing alternatives. This looming wave of novice online freelancers, who likely intend to remain for the long term, poses many questions about successful strategies for \red{online} freelancing. 

\red{In a systematic review of the sharing economy in computing research, Dillahunt et. al. focused on the computing (HCI in particular) community's contributions toward the sharing economy as well as underexplored and unexplored topics for future research. \cite{Dillahunt_Wang_Wheeler_Cheng_Hecht_Zhu_2017}.} 
\red{ This prior literature review suggests that existing HCI studies on the sharing economy has been largely descriptive and qualitative.
\final{To diversify the range of HCI approaches applied toward the gig work context, we present here a quantitative study that leverages the aggregate past experiences of workers.} The literature review also suggests a need for human-centered optimization that increases the decisional autonomy and long-term performance of workers, while minimizing overheads such as reduced availabilities, monetary cost, and worker burnout \cite{Dillahunt_Wang_Wheeler_Cheng_Hecht_Zhu_2017}.} Within this empirical investigation we offer an initial identification of effective \red{project-level and long-term} self-management strategies, to inform novice freelancers about what, when, and how much to communicate with their clients, and more generally about the overall marginal benefit (or cost) of `personalizing' service delivery, e.g., tailoring communications or work schedules to one's client. 

\subsection{\final{Strategies in Online Freelancing}}

Compared to the more organization-centered employees of the traditional labor market, workers of the gig economy are individually-organized and experience many of the same challenges as entrepreneurs at the beginning of their gig career. With this in mind, we draw on literature in entrepreneurship to identify relevant strategies that could be applied to online freelancing. \red{For resource-constrained entrepreneurs, prior work} \cite{Busch2021-ls, Vanevenhoven2011-bx, Atarah2021-ox} \red{ identified the strategy of } bricolage - the act of creatively working with available, limited resources, and adapting them toward new or important purposes; or as Levi-Strauss put it: make do with ``whatever is at hand" \cite{lvi1966savage}. Online freelancers \red{are} also bricoleurs \red{when they leverage} available resources (such as messaging systems, \red{client reviews} and job descriptions to learn more about the requirements of a gig) to tailor their pitches to\red{ward} employers accordingly. 

One particular way of circumventing resource constraints is by engaging in network bricolage - where workers utilize their network resources in a manner \red{that is} different from the original basis for the connection, thereby creating new opportunities \cite{Chang2019-ri}. In the online gig economy, such resource-creators may find relational support by engaging with offline networks, developing a mentorship relationship with senior freelancers, or cold-emailing potential clients. Because the ``infrastructure supporting individuals' careers in the gig economy is deeply relational in nature" \cite{Ashford2018-dw}, it is important that freelancers accrue portable human capital. After amassing such social capital, workers may maintain \red{their reputation} using various strategies such as keeping a \red{high and} positive \red{rating}, reaching out to past clients or cultivating relational agility by productively forming, maintaining and dissolving work relationships \cite{Sutherland2020-wk, Blaising2021-bn}. 

\red{To overcome information asymmetries in online labor markets, workers may engage in prosocial network bricolage behaviors to build connections, so that successful freelancers can share experiential knowledge and novice ones can gain from the collective advice of more veteran peers. Social media groups, for instance, serves as a key resource for informational peer support for rideshare drivers, helping them alleviate the burdens of atomization of being geographically dispersed \cite{Yao_Weden_Emerlyn_Zhu_Kraut_2021}. In freelancing platforms such as Upwork, novice workers may leverage the advice of more senior and successful freelancers \cite{platform_manage}. To circumvent power asymmetries and platformic management, workers might take courses to gain algorithmic literacy about the platform \cite{platform_manage} or experiment with it themselves to develop strategies such as ``saving searches" to improve recommended jobs or asking clients to report multiple hours of work as one condensed hour to improve their hourly rates \cite{Jarrahi_Sutherland_2019}.}  
    
\red{For digitally-mediated work, many} of these network bricolage behaviors involve some means of direct communication between the client and worker, such as emailing or messaging. \red{While there are workers who use opt to use external communication tools to deal with technical inefficiencies, unreliability and monitoring concerns \cite{platform_manage, Jarrahi_Sutherland_2019}, } most online labor platforms provide a form of direct messaging system to mediate textual exchanges between workers and employers\red{. In fact, one of the platform's core functionalities is to facilitate communication between transacting parties \cite{platform_manage}.} \red{Thus, we focus} primarily on communication strategies that freelancers commonly engage in when \red{chatting with} potential clients. 

\subsection{\final{Stages of the Project Lifecycle}}

While we intend to study strategies that are applicable during all periods of a freelancer's career, it is important to \red{distinguish} different points of a project cycle. In this paper, we will consider both the initial, pre-contract (bidding) stage and the project execution stage:

\begin{enumerate}
    \item Bidding stage: Client may interact with multiple freelancers prior to offering the job to selected candidate(s), both clients and freelancers may negotiate and clarify the scope of work before finalizing on price
    \item Execution stage: After the client makes the job offer and sends deposits promised compensation via an Escrow, the worker begins work to complete job demands 
\end{enumerate}

At the initial bidding stage (1), freelancers may attempt different techniques to garner the attention of a potential employer. These may include stylistic techniques such as the use of custom signatures and uppercase words for emphasis, as well as content curation strategies such as using templates to quickly initiate conversations with multiple employers. On the other spectrum, some freelancers may also choose to personalize the content of their (bid) messages to accommodate job demands. Following initial introductions, the freelancer and client negotiate to settle \red{on a} price and review contract terms to clear up points of confusion. At this point, if the project is ill-matched, either party may choose to reject the collaboration. During the execution stage (2), the worker may provide progress updates, request additional clarifications, or ask for milestone payments while the client can ask for updates to monitor progress. Note that after the successful completion (or abandonment) of the project, users may request reviews, provide reminders about payments, or bring up opportunities for future collaborations. 

Evidence \red{abounds that} direct communication \red{benefits sellers on} digital marketplaces, including Alibaba, Amazon, and Travelweb \cite{trust, alibaba}. For instance, the use of live chat on Alibaba can increase purchase probability of tablets by 15.99\% \cite{alibaba}. \red{In online freelancing}, it has been found that \red{workers} are 8.9\% more likely to be hired if they initiate a direct message to a potential employer when submitting their bid \cite{Hong2021-hj}. \red{In our own sample data, we qualitatively observed comparable patterns of benefit}. Specifically, we see that freelancers who employ high-quality templates (e.g., containing examples of past work and self-promotional messages) tend to receive more responses from clients. Further, we observe that freelancers who proactively provide progress updates to their clients during the course of a project are more likely to successfully complete the work and receive payment.  

With the exception of \cite{Hong2021-hj}, the present body of literature has yet to systematically investigate the communication strategies employed by workers in the context of online labor platforms. We thus currently have a limited understanding of the alternative work strategies that gig workers employ during job search and project execution, and the relative efficacy of each. Hence, we explore those questions here.

\section{Standardization vs Personalization in Communication} \label{grounding}

At a high-level, online freelancers can be expected to adopt two main strategies: they can \textbf{\textit{standardize}} their messaging practices (using techniques such as templated content or regular messaging hours) or \textbf{\textit{personalize}} their communication to cater toward desires of an employer (by messaging during the client's preferred hours or curating their proposals to fit the needs of a job). While standardization offers efficiency gains by saving time and effort for freelancers, personalization can facilitate smoother correspondence with employers by providing them just the information they need, when they need it. However it remains unclear whether it's more efficient \red{and beneficial} for workers to take the standardized approach of offering their services to a large group of clients, or focus on more personalized services that accommodates the individual needs of each job and client. 

\red{A past study on telephone surveys examined the tension between standardization requirements (interviewers are prohibited from laughing during survey administration to maintain consistency across surveys) and rapport-maintenance expectations, which can manifest when survey respondents initiate a laughter invitation \cite{rapport}. Although the interviewers of this study declined to join in on respondents' laughter invitations, there was no exploration of whether the breakage or maintenance of rapport through (the lack of) laughter responses affects the quality of the surveys \final{-- a success measure that would have been appropriate for this laughter study}. In this piece we endeavor to explore \final{how the tradeoff between standardization and personalization communication techniques affect success outcomes such as job acquisition and completion.} }

\red{However we acknowledge that the two are not mutually exclusive practices -- a freelancer may choose their strategies depending on plethora factors such as their familiarity with the client, expertise with the job posting, the stage of the job cycle they're currently in, or their personal bandwidth and availability. Standardization and personalization may also be exhibited in a different ways -- freelancers may remain temporally consistent in their responses to client requests while remaining delivering standardized, templated message response content. So in addition to tradeoffs, we plan to also investigate how these strategies can interact with other factors and exhibit different effects when applied to multiple contexts.}

\subsection{Standardization}

According to De Vries, standardardization is defined as \textit{``the activity of establishing and recording a limited set of solutions to ... problems directed at benefits for the party or parties involved balancing their needs ... expecting that these solutions will be repeatedly or continuously used during a certain period by a substantial number of parties for whom they are meant"} \cite{Munstermann2008-yj}. \red{For independent contractors, it is certainly expected that their services will be used among multiple parties.} Meanwhile Lehr \cite{Hanseth1996-ux} considers standardization to be the \textit{``social and technical process of developing the underlying artifact related to [information infrastructure] - ... standards that govern the communicative patterns"}. \red{Such procedures for developing standardized communication brings us closer to process standardization, involving ``the development of a common approach to such activities as establishing (and evaluating) a distributor network ... the underlying approach to relationship development strategies"\cite{Griffith}. In \cite{Griffith}, Griffith et. al. explores communication strategies applied across different cultural contexts, and finds that standard processes may beneficial when applied to nations of similar cultural types, but not necessarily on a global scale. But to the extent of our knowledge, there exists no prior establishment of standardization measures for communicative practices in global, online freelancing platforms.}
 
\red{In corporate contexts,} communication has been found to benefit organizations during challenging or exciting times, while ill-conceived or incomplete communication caused by poorly constructed and delayed messages has the potential to turn small issues into major crises \cite{Newman2016-nd}. In business contexts (service sectors \red{in particular}), process standardization offers profitable outcomes by helping define clear and precise output objectives for the service provider, and by better facilitating communication and coordination between exchange partners through increased uniformity of process activities \cite{Ramakumar2004-fv, Munstermann2010-yf, Wuellenweber2009-pa}. \red{For the freelancing context, we consider content standardization to be the process where workers repeatedly use messages constructed from templates to promote their services toward multiple clients or job postings, and temporal standardization to be the practice where freelancers message around a fixed time of the day across various projects.}

In May 2020, Upwork (a leading online freelancing platform) presented a set of proposal templates as resources to guide beginner freelancers\red{. B}ut by June \red{of} 2021, the use of templates is no longer recommended and instead it is suggested that freelancers should \textit{``focus more on [specific] project needs}", \red{suggesting that the benefits and harms of template use in bidding could be complicated} \cite{noauthor_undated-wy}.

\subsection{Personalization}
\red{There's no shortage of existing frameworks for personalization, especially within marketing literature and persuasive  (mobile \& e-commerce) technologies \cite{fan2006personalization, zhang2007toward}. For technology, Blom defined personalization as ``a process that changes the functionality, interface, information content, or distinctiveness of a system to increase its personal relevance to an individual" \cite{Blom_2000}; meanwhile the Personalization Consortium defined it as ``the use of technology and customer information to tailor electronic commerce interactions between a business and each individual customer." \cite{consortium}. In business contexts, personalization may entail ``Customizing some feature of a product or service so that the customer enjoys more convenience, lower cost, or some other benefit" \cite{peppers}, and in internet marketing it has been considered ``A specialized form of product differentiation, in which a solution is tailored for a specific individual" \cite{hanson2000principles}.}

\red{The definitions of personalization presented so far addresses the specialization of content or a service for an individual, which can be achieved through the presentation of curated options based on known information about their target user or customer - a process that Churchull refers to as \textit{outcome personalization} \cite{Churchill_2013}. But consider \textit{process personalization} (which occurs in service encounters) where information is collected about a customer through realtime interactions, and instead of focusing solely on the outcomes, increasing the quality of interaction and delivery are also a part of the objective \cite{Churchill_2013}. Reflecting this more interactive definition, the Personalization Consortium expands on their previous definition: ``Using information either previously obtained or provided in real-time about the customer, the exchange between the parties is altered to fit that customer's stated needs as well as needs perceived by the business based on the available customer information" \cite{consortium}. Dyche and Robert's respective definitions are also more process-oriented ``the capability to customize customer communication based on knowledge preferences and behaviors at the time of interaction" \cite{roberts2012internet}, and ``The process of preparing an individualized communication for a specific person based on stated or implied preferences" \cite{dyche2002crm}. }

\red{In online chats that are devoid of physical signals from body language or tone of voice, personalizing interactions through messaging content and pace can be of paramount importance for improving interactions between transacting parties. Indeed, Blom identified that a key motivation for using personalization to be the enablement of access to information content [for the customer/client], which can help facilitate interactions and transactions \cite{Blom_2000}. Process personalization can also be personalized\final{:} ``customized personalization is about personalizing to the consumer's interactional style and needs in the moment, as well as more stable or longer-term facets such as their demographic profile and/or manifest tastes", but the customized personalization can have varying effects depending on context and degree (obsequiousness, for instance, can be upsetting) \cite{Churchill_2013}. In service encounters, personalization improves customer impressions \cite{Surprenant1987-sz} and in a persuasion study, personalized mobile messages successfully helped individuals by significantly reducing daily snacking \cite{Kaptein_Markopoulos_Ruyter_Aarts_2015}. But in the context of student-advisor instant messaging interactions \cite{Siqueira2009-md}, the adoption of accommodating temporal patterns has been shown to disrupt one's own temporal consistency. }

For the freelancing context, we define personalization as the way in which a worker caters to the needs of a client by incorporating relevant job specifications into their message text or by client messages, at the expense of their own work schedule or time zone. At the outset, the relationship between standardization and personalization may seem divergent and potentially conflicting, and prior literature has long recognized the tension between information standardization and flexibility \cite{Hanseth1996-ux}. However, we discuss below how these practices might coexist and the potential trade-offs between the two in terms of their effects on outcome success during different stages of a project, as well as over a freelancer's long term career trajectory.

\subsection{\final{Communication Strategies Across Stages}}
\subsubsection{Bidding phase strategies}
Vetting for a job in online freelancing platforms may seem intimidating to many workers, especially to beginners who may be submitting their first few bids. But as \cite{Hong2021-hj} \final{has} show\red{n}, reaching out to clients has a significant and positive impact on a freelancer's chances of procuring a job. Among the workers who do initiate conversations with clients, we consider whether content curation would have an effect on hiring probabilities. At the beginning of this investigation, the use of proposal (bid) templates was still a recommended practice by platforms such as Upwork. Since sending out templated first messages to multiple clients en masse can save time and maximize resource utility, we expected freelancers to leverage the advantages of template use when initiating conversations with clients. 

Since the online gig economy is structured as a reverse auction market, clients are often subject to information asymmetries. In particular, the lack of insight into worker bandwidth may lead to wasted time and effort for the client \cite{Horton2018-eq}. Receiving direct messages from freelancers can help clients overcome such obscurity since the gesture of outreach serves as an indicator for clients to gauge the bandwidth and capacities of a freelancer. While we know that outreach in general has a positive effect on hiring probabilities \cite{Hong2021-hj}, we may expect templated messages to induce the opposite effect: clients might observe that the freelancer has the time, capacity and perhaps even desperation \cite{dunn2020motivation} to find work, but not the resources \red{necessary} to personalize the content of their message to target the needs of their individual project. Hence, we can expect clients to hire more freelancers who demonstrate sincerity through individualized content curation in their first outreach message and bid proposal texts:

\textbf{Hypothesis 1:} During the bidding stage of a project, we posit that
\begin{enumerate} [label=(\alph*)]
    \item Standardizing first message text will \textit{decrease} the probability of winning the bid.
    \item Personalizing bid text to match job description requirements will \textit{increase} the probability of winning the bid.
\end{enumerate}

\subsubsection{Execution phase strategies}
Due to intense competition in the online labor market \cite{dunn2017digital}, freelancers may feel pressured to respond to client requests as quickly as possible to minimize the chances of the clients noticing and hiring other competitors. However, this may reduce productivity during the execution phase since ``constantly attending to IM ... may prevent users from performing tasks efficiently, leaving them frustrated." \cite{Avrahami2004-cd}. Furthermore, the cognitive switching costs accrued by toggling between attending to messages and focusing on work is especially pronounced during the execution stage: ``the time to switch to the message was significantly slower when the notification arrived during the execution phase than either other phase" \cite{Cutrell2000-ap}. 

The expectation to remain responsive may disrupt freelancers' workflows, allowing clients to interrupt them when completing a task, \red{thereby} reducing their working efficiency. Some direct messages may exhibit characteristics of outeraction - communicative processes people use to connect with each other and to manage communication, rather than to information exchange. Outeractions can be especially disruptive because the content of the exchange is unrelated to the freelancer's task at hand: ``time spent on messages and time to resume the search task were both longer when the message was irrelevant than when it was relevant" \cite{Cutrell2000-ap}. Hence, our second hypothesis examines how personalizing practices during the execution stage, such as responsiveness and accommodating the ``regular hours" of a client, can affect project completion outcomes:

\textbf{Hypothesis 2:} During the execution stage of a project, we expect that
\begin{enumerate} [label=(\alph*)]
    \item Responding during a standardized period during the day will \textit{improve} the probability of completing a project.
    \item Personalizing response times (increasing responsiveness) will \textit{negatively impact} the probability of completing a project. 
\end{enumerate}

\subsubsection{Messaging techniques and revenue}

Outcomes such as award and completion statuses serve to measure the success of various messaging practices at the individual project level. However, to evaluate the impact of these practices over the long term, we must observe a more aggregated measure of the freelancer such as their monthly revenue or earning efficiency. \red{With the exception of a study that found multitasking among Turkers to generate higher income more quickly \cite{Brewster_Fitzpatrick_Cox_Kostakos_Lascau_Gould_Cox_Karmannaya_Brumby_2019}, there's a scarcity of literature available investigating the effects of messaging patterns on freelancer revenue.} 

\red{We think there} is reason to believe that over the long term, standardization can help \red{freelancers generate} revenue while personalization will hurt \red{their quantity of earnings} because \red{personalizing content for each specific client} and always \red{being available for and} responsive to clients can be draining and unsustainable over the long term. \red{But on the other hand}, the opposite might also hold true: freelancers could adapt to manage their time in a way that they personalize and thrive for each of their projects without experiencing burnout. Thus, we leave the effects of standardization and \red{personalization} on revenue as a research question to be examined:

\textbf{Research question 3:} how do standardizing and personalizing help or harm revenue?

\section{Research Context and Methods}
\subsection{Study Platform}
To conduct this study, we obtained data from a corporate partner \red{(whose specific name will not be disclosed per agreements for data sharing)} that is a leading platform in online freelancing. \final{Example categories of work include data entry, software development, design, writing, etc.} The dataset we acquired consisted of \final{2,031,068} messages, from \final{56,222 } projects posted between January 1, 2010 and March 1, 2010, \final{ involving 58,397 freelancers and 25,480} clients. For each project we observed their associated project descriptions, bid text, messages, as well as timestamps for these artifacts such as the submission and award dates of bids, the completion and payment times, as well as individual message timestamps. We did not impose limitations based on project category. For each stage of a project, we constructed two \red{separate data frames using this sample}. Observations in the first \red{frame} consisted of worker-job pairs (or conversations) that incorporated worker-related information such as bid price, bid text, reputation as well as information associated with the job, including project description text, submit date and buyer identification. In a separate freelancer\red{-level} \red{frame} we included long term worker-related attributes such as average bidding price and bid volumes.

\subsection{Measures of Key Variables}

We operationalized standardization and personalization in communication depending on the phase of the project. To more precisely capture standardization in the execution phase, we removed freelancers who multitask and work on more than one project at once -- multitaskers represented roughly 12\% of those who were awarded projects.

\subsubsection{Bidding phase strategies}  
  \setlength\itemsep{1em}

\begin{itemize}
\item \textbf{First Message Standardization:} To measure the extent to which freelancers \textit{standardize} content in a conversation (i.e. worker-job pair) during the initial bidding stage, we calculated the \textit{first-message similarity}. We obtain this measure for a particular conversation by calculating the cosine distance between the freelancer's vectorized \footnote{ The vectorization approach we use is to simply create counters for word occurrences in the messages.} first message in the current project and the vectorized first message of their most recent prior project. Hence, freelancers who use the same set of words across first messages to multiple clients tend to score higher in this measure since they are more likely to employ standardized templates when conducting outreach.

\item[\ding{46}] \red{\underline{Example:} If freelancer $F$ uses a template $T$ and sends $T$ in their first message to the clients in both projects $P_1$ and $P_2$ (assuming $P_2$ immediately follows $P_1$), they will receive a measure of 1 for their first message standardization for project $P_2$. But if for their project $P_3$, $F$ sends a  first message that is completely different to the previous two (i.e. no words in the first message of $P_3$ matches those in $T$), then the standardization measure for $P_3$ would be 0.} Since this measure only concerns the first message content sent by the freelancer in each project, it will only be used as an explanatory variable for the bidding stage model.

\item  \textbf{Bid Text Personalization:} To quantify the amount of \textit{personalization} that freelancers employ in the bidding stage, we computed the level of curation in the freelancer's bid text. This measure represents the degree of likeness (again obtained via cosine similarity) between the textual content of a freelancer's bid application and its associated project description post (submitted by the potential employer). Accordingly, freelancers who choose to include words and mirror content from the client's job post are considered to have higher measures of personalization. 
\item[\ding{46}] \red{\underline{Example:} Freelancer $F$ submits a bid application $B_1$ to project $P_1$. $B_1$ borrows many words from the job posting. Subsequently, $F$ applies to another project $P_2$ with $B_2$, but $B_2$ did not make use of any text from the job description. Freelancer $F$ would have a higher measure of content personalization for $P_1$ than for $P_2$.} Similar to first-message similarity, this variable measures a practice that can only be executed in the bidding stage, and will therefore only be used as a predictor variable for hiring outcomes.
\end{itemize}
\begin{table*}[h]
\begin{tabular}{lcccccccc}
\hline
\textbf{Bidding stage (project-level \red{frame})} 
& \multicolumn{1}{l}{}                      
& \multicolumn{1}{l}{}                                   
& \multicolumn{1}{l}{}       
& \multicolumn{1}{l}{}       
& \multicolumn{1}{l}{}      
& \multicolumn{1}{l}{} 
& \multicolumn{1}{l}{}
& \multicolumn{1}{l}{} \\ \hline
{Measure}                  
& \multicolumn{1}{l} {Mean} & \multicolumn{1}{l}{{Standard Deviation}} & \multicolumn{5}{c}{Correlations} \\ \cline{4-9} 
& \multicolumn{1}{l}{}                      
& \multicolumn{1}{l}{}                                    
& 1                          
& 2                          
& 3                         
& 4                    
& 5
& 6\\ 
\hline
\textit{1. First message standardization} & \ 
.49\red{4}                                    
& \red{.344}                                               
& \multicolumn{1}{l}{}       & \multicolumn{1}{l}{}       
& \multicolumn{1}{l}{}      & \multicolumn{1}{l}{} 
& \multicolumn{1}{l}{} & \multicolumn{1}{l}{}\\
\textit{2. Bid text personalization}      
& .128                                   
& \red{.124}                                               
& \multicolumn{1}{l}{-.106} & \multicolumn{1}{l}{}       
& \multicolumn{1}{l}{}      & \multicolumn{1}{l}{} 
& \multicolumn{1}{l}{} & \multicolumn{1}{l}{}\\
\textit{3. Had prior reviews}             
& .61\red{1}                                   
& .487                                               
& \multicolumn{1}{l}{.211}  & \multicolumn{1}{l}{-.175} 
& \multicolumn{1}{l}{}      & \multicolumn{1}{l}{} 
& \multicolumn{1}{l}{} & \multicolumn{1}{l}{} \\
\textit{4. Bid price (log)}               
& 5.05                                     
& 1.36                                               
& \multicolumn{1}{l}{.121}  & \multicolumn{1}{l}{.031}  
& \multicolumn{1}{l}{.043} & \multicolumn{1}{l}{} 
& \multicolumn{1}{l}{} & \multicolumn{1}{l}{} \\
\textit{5. Bids won (log)}               
& 2.12
& 1.93                                              
& \multicolumn{1}{l}{\red{.212}}  & \multicolumn{1}{l}{\red{-.154}}  
& \multicolumn{1}{l}{\red{.758}} & \multicolumn{1}{l}{\red{.078}} 
& \multicolumn{1}{l}{} & \multicolumn{1}{l}{} \\ \hline

\textbf{Execution stage (project-level \red{frame})}
& \multicolumn{2}{c}{}       & \multicolumn{1}{l}{}       
& \multicolumn{1}{l}{}       & \multicolumn{1}{l}{}      
& \multicolumn{1}{l}{}       & \multicolumn{1}{l}{} \\ \hline

\textit{1. Response time standardization} 
& 7.05e04 & 4.15e03 & & & & &                      \\
\textit{2. Response time personalization} 
& \red{5.22}e5 & 6.23e05 & -.025 & & & &                      \\
\textit{3. Had prior reviews} 
& .612 & .487 & -.281 & .008 & & &                      \\
\textit{4. Bid price (log)}                        
& 5.06 & 1.36 & .050 & -.021 & 
.043 & &                     \\
\textit{5. Freelancer message count}      & 4.82 & 7.90 & -.082                     & .002                      & .107                     & -.018               & \red{}             \\
\textit{6. Projects completed (log)}
& \red{4.35} & \red{1.38} & \red{-.107} 
& \red{-.005} & \red{.251} & \red{.029} & \red{.040} 
\\ \hline

\textbf{Freelancer - level \red{frame}}                  
& \multicolumn{1}{l}{} & \multicolumn{1}{l}{}
& \multicolumn{1}{l}{}       & \multicolumn{1}{l}{}       & \multicolumn{1}{l}{}      & \multicolumn{1}{l}{} & \multicolumn{1}{l}{} \\ \hline
\textit{1. Average first message standardization}  & \red{.472}                                    & .253                                               &                            &                            &                           &                      &                      \\
\textit{2. Average bid text personalization}       & .13\red{5}                                   & \red{.100}                                              & -.\red{189}                     &                            &                           &                      &                      \\
\textit{3. Average response time standardization}  & 7.03e04
& 3.9\red{8}e03                   
& -.0\red{75}
& .1\red{59}                &                           &                      &                      \\
\textit{4. Average response time personalization}  & \red{5.69e5}                                     & \red{3.72e04}
& -.01\red{8}                     
& -.0\red{77}                
& -.\red{129}                    &                      &                      \\
\textit{5. Average bid price (log)}       
& 4.\red{86}                                     
& .8\red{96}                                            
& .1\red{66}                     
& -.0\red{36}                 
& .00\red{8}               
& -.0\red{66}
&                      \\
\textit{6. Had prior reviews average}              & .63\red{5}                                    & .4\red{52}                                              & .2\red{44}                      & -.2\red{56}                     & -.237                    & \red{.119}               & .12\red{0}               
\\ \hline
\end{tabular}
\caption{ Correlations, means and standard deviations of explanatory variables}
\end{table*}

\subsubsection{Execution phase strategies}

\begin{itemize}
\setlength\itemsep{1em}
    \item \textbf{Response Time Standardization: }After a freelancer makes it past the selection stage and is awarded the job offer, we look at the effects of qualities such as timing on a freelancer's likelihood of successfully completing a project. In particular, we measure \textit{standardization} in this stage by computing  the schedule regularity of a freelancer within a particular project. 
    \red{To compute this measure, we first find the standard deviation in the timing of the day for a freelancer's messages across all their projects (this is a freelancer-level measure). But since that measures the variance in schedules, we invert the standard deviation by subtracting it from the total number of seconds in a day to better represent schedule regularity. }
    
    \item[\ding{46}]\red{\underline{High regularity example:} Freelancer $F$ sent a total of two messages, one at 11:02am and another at 11:12am \footnote{\red{Note that the day when the messages were sent does not affect this variable as it measures regularity on a daily basis.}}. The standard deviation of $F$'s messages is five minutes, which means that the measure of schedule regularity is quite strong at 23 hours and 55 minutes.} 
    
    \item[\ding{46}] \red{\underline{Low regularity example:} By contrast, freelancer $G$ sent two messages that are much further apart in the day - one of them at midnight (00:00:00) and another at noon (12:00:00). The standard deviation of of $G$'s messages is six hours, and their schedule regularity is much lower (at 18 hours).} Thus, the smaller the deviation in message sending times, \red{the less likely that the freelancer compromises their own routines to accommodate clients' timezones or schedules. }

    \item\textbf{Response Time Personalization:} To estimate whether \textit{personalization} affects the likelihood of project completion, we calculated for each freelancer-project pair its \textit{responsiveness}. 
    \red{First we determine the response gap of a message by calculating the amount of time it takes for a freelancer to respond to a message sent by the client}\footnote{ If an employer sends multiple messages before receiving a response, we consider the response time to be the difference in time between the freelancer's first response and the employer's \textit{first} message that has not yet received a response.}. 
    \red{Then all we average these response gaps across all messages of the conversation to obtain an aggregated measure at the worker-project level. Once again, we invert this measure by subtracting it from the the total number of seconds in a week so that it embodies responsiveness instead of response times.}
    
    \item[\ding{46}] \red{\underline{High responsiveness example: } Freelancer $F$ responded to two client messages in project $P$. For the first message they replied back 90 minutes after the client's message while the second response took them 30 minutes. The average response time of freelancer $F$ in project $P$ is very quick at 1 hour, which means that $F$'s average responsiveness in project $P$ is 6 days and 23 hours.}
    
    \item[\ding{46}] \red{\underline{Low responsiveness example: } Now let's say freelancer $G$ also worked on project $P$, and responded to two client messages for this project as well. Their first reply only took 1 hour but they missed the client's second message and ended up taking 9 days and 23 hours (a total of 239 hours) to respond. So the average response time for freelancer $G$ in project $P$ is much slower (at 10 days, or 240 hours), implying that their responsiveness for project $P$ is much lower at 4 days.}\red{ Intuitively, freelancers incur shorter gaps when they are being more responsive}, which also demonstrates greater amounts of personalization in terms of message timing.
    
\end{itemize}

\subsection{Outcome Measures and Control Variables}

For measuring success at different stages, we gather the \textbf{job award status} to assess the outcome of the bidding stage, \textbf{project completion status} for the execution stage, and \textbf{overall monthly revenue} to account for long-term earnings. Both award and completion status are binary variables where ``awarded" or ``complete" corresponds with 1 while all other statuses (``rejected", ``incomplete", or ``pending") are marked as \red{0}. Revenue is a dollar amount calculated on a monthly basis, the final value of of revenue per month is normalized with standard scaling.

Beyond these key explanatory variables, we also include other controlled variables: \red{reputation is measured by whether the worker has received reviews in the past (\textit{had prior reviews}), \textit{bid price} is the amount that the freelancer is proposing to charge for their work (this variable is log-transformed to remove skewedness), \textit{freelancer message count} is the total number of messages the freelancer sends within the project, number of bids won and projects completed account for how many projects the freelancer's has historically being hired for and completed, respectively, and are also log-transformed. All predictor variables are normalized for analysis via standard scaling. In Table 1 we provide descriptive statistics of both key and control variables for each of our models.}

\subsection{Statistical Models }
Using separate linear regression models for different stages, we observe the effects of standardization and personalization techniques toward project hire, completion outcomes as well as earnings. When testing the hypotheses about the bidding and execution stages (H1 and H2), we eliminate the possibility that hiring and completion statuses are jointly determined with our explanatory variable by including a project-level fixed effects when running the logistic regression model. This captures time-invariant and job specific properties that might impact the model outcomes, as well as employer-level fixed effects, since there can be only one employer per job. 

The models also include observable worker characteristics that may vary across bids such as reputation status and bid price. At the revenue level (R3), we first ran a regression model that used the four aforementioned strategies \red{(measured by our key variables)} to predict monthly revenue. Subsequently, we used the two bidding stage measures to predict the total freelancer bidding volume over the three month period to provide further insights for results of the revenue model.

\section{Results}
\subsection{\final{Bidding Strategies' Impacts on Hiring}}
\begin{table}[H]
\begin{tabular}{l c c}
\hline
\begin{tabular}[c]{@{}l@{}}Dependent Variable:\\ \textbf{Job award}\end{tabular} & Coefficient   & Standard Error  \\ \hline
\textit{First message standardization}                                            & -.0\red{30}1*** & 8.2\red{3}e-05   \\ \hline
\textit{Bid text personalization}                                                            & .03\red{58}***  & 2.30e-03   \\ \hline
\textit{Had prior reviews}                                                            & \red{3.90e-03}***  & \red{7.72e-04}     \\ \hline
\textit{Bid price (log)}                                                      & -.02\red{58}*** & 5.\red{36}e-04   \\ \hline
\red{\textit{No. bids won (log)}}                                                      & \red{7.30e-03***} & \red{2.38e-04}    \\ \hline
Number of observations                                                  & \multicolumn{2}{c}{603,286}    \\ \hline
\multicolumn{3}{c}{*** signifies a p-value \textless .001, errors are clustered by project}           \\ \hline
\end{tabular}
\caption{{ Bidding stage regression model with project fixed effects predicting job awards. }}
\end{table}

Table \red{2} shows the bidding stage regression results, where we explore the impacts of standardizing and personalizing first messages on the project award outcome (1 is awarded and 0 if not). The coefficients show that increasing standardization during the bidding stage hurts a freelancer's hiring probabilities, \red{thereby supporting H1a. Specifically, standardizing first message content by one standard of deviation reduces their winning probabilities by .03\%. Meanwhile, personalizing and curating the contents of a bid proposal based on the job posts increases their chances of winning the project (which is in alignment with H1b), but only slightly -- personalizing bids by one standard of deviation improves award probability by .036\%. } 

Note that we also controlled for freelancers' bid prices (which were log transformed after adding one since the log of zero is undefined),  reputation -- measured via the dummy variable \textit{had prior reviews}, which represents whether the freelancer has received a rating for their work in the past, and a historical bidding success variable -- the number of bids the freelancer won prior to the current project. We intentionally chose to not include actual rating values because the majority of ratings are positive and most jobs do not end up receiving reviews - their inclusion would cause an inflated measure of reputation. As one would expect, \red{having previously won bids and} reviews \red{to showcase on the profile} is favorable for hiring, whereas bidding at a higher price harms hiring probabilities \red{of a freelancer}.

\subsection{\final{Execution Strategies' Effects on Completion}}
\begin{table}[h]
\begin{tabular}{lcc}
\hline
\begin{tabular}[c]{@{}l@{}}Dependent Variable:\\ \textbf{Job completion}\end{tabular} 

& Coefficient   & Standard Error \\ \hline
\textit{Response time standardization}                                                             & \red{1.15e-07}  & 1.1\red{6}e-07 \\ \hline
\textit{Response time personalization}                                                                   & -9.\red{10}e-0\red{9}*** & \red{1.63}e-09  \\ \hline
\textit{Had prior reviews}                                                                    & \red{5.90}e-03***  & \red{2.01}e-03 \\ \hline
\textit{Bid price (log)}                                                              & -1.2\red{7}e-02*** & 1.0\red{3}e-03 \\ \hline
\textit{Freelancer message count}                                                        & 8.\red{1}6e-03***  & 5.\red{52}e-04\\ \hline
\red{\textit{No. completed projects} }
& \red{1.36e-03 .}  & \red{7.33e-04} \\ \hline
Number of observations                                                          & \multicolumn{2}{c }{1\red{10,797}}    \\ \hline
\multicolumn{3}{ c }{*** signifies a p-value \textless .001 and . denotes a p-value < .1}
 \\ \multicolumn{3}{ c }{Errors are again clustered by project} \\ \hline
\end{tabular}
\caption{Execution stage regression model with project fixed effects predicting
\label{execution_table}
job completions.}

\end{table}
Table \red{3} shows our results for the execution stage model, where we explore the impacts of standardizing or personalizing responses time on the job completion.
In this stage, we observe that \red{ in alignment with H1a, being online at regular hours of the day has a small and positive but insignificant effect on a} freelancer's chances of completing a project
Meanwhile, being highly responsive to client messages (the personalization technique) significantly hurts completion, \red{which is in agreement with H2b, but the effect is negligible}. 

Reputation and bid prices have a similar effect as in the bidding stage model. This suggests that reputable freelancers have higher chances of satisfying the demands of a client. Workers who demand higher payments will have a harder time gaining approval from their clients, since more costly payments \red{will likely} lead to increased expectations for work quality. \red{Having received ratings for prior work is positively correlated as well.} We also controlled for the number of messages that a freelancer sends within the project, since \red{message frequency} will have a consequential impact on the variance/regularity of a worker's messaging schedule, \red{and found that messaging more positively impacts completion probabilities. Meanwhile, having successfully completed projects slightly helps execution of the current one.}

\subsection{\final{Long-term Strategies' Impact on Earnings}}
Our earnings model uses a freelancer-level instead of a project-level \red{frame} to capture revenue from all jobs of a month. Here we measure the effects of the same two pairs of standardization and personalization techniques above \red{to investigate the question posed in R3}. The four measurements are aggregated for each freelancer \red{frame} by averaging, and fixed effects are added to account for time variance. 

Table \red{4} reveals that only messaging practices in the bidding stage had significant impacts on overall revenue. Specifically, standardization has a positive effect on revenue \red{- increasing content standardization by one standard of deviation results in a .12\% growth in monthly revenue,} likely because it enables workers to submit more bids. \red{Meanwhile} bid personalization no longer offers the same enhancing effects it had at the project level. \red{In fact, personalizing bid content by one standard of deviation can cost workers .15\% of their monthly revenue}. Reputation continues to impact success in the same ways as before, and bid higher for individual projects naturally increases overall freelancer earnings.

\begin{table}[h]
\begin{tabular}{l c c}
\hline
\begin{tabular}[c]{@{}l@{}}Dependent Variable:\\ \textbf{Monthly revenue}\end{tabular}  
& \multicolumn{1}{l }{Coefficient}          & \multicolumn{1}{l }{\final{Standard Err.}}  \\ \hline
\textit{\final{Avg. first msg. standardization}}                                           & \red{.123}*** & \red{3.23}e-02                     \\ \hline
\textit{\final{Avg. bid text personalization}}                                       & -.1\red{46}** & 3.\red{6}2e-02     \\ \hline
\textit{\final{Avg. response time standardization}}                                  & -\red{1.18e-06} & 1.\red{37}e-06 \\ \hline
\textit{\final{Avg. response time personalization}}                                  & -\red{1.52e-07} & \red{1.61e-07}    \\ \hline
\textit{Avg. bid price}                                                      & .2\red{17}*** & 1.\red{22}e-02  \\ \hline
\textit{Had prior reviews average} 
& .36\red{0}*** & 1.\red{61}e-02    \\ \hline
\textit{Number of observations}                                                 & \multicolumn{2}{c}{161\red{49}}                                                               \\ \hline
\multicolumn{3}{ c }{\begin{tabular}[c]{@{}c@{}}\text{*** signifies a p-value \textless .001, ** denotes a p-value \textless .01}\\ \text{Errors are clustered by month}\end{tabular}} \\ \hline
\end{tabular}
\caption{ Freelancer-level regression predicting monthly revenue with monthly fixed effects.}
\end{table}

\red{To test our hypothesis that the inverted effects of content standardization is related to how} it enables workers to take on larger volumes of work, we ran an additional model using bid volume as the dependent variable. \red{The results show that increasing first message standardization by one standard of deviation} during bidding \red{can allow workers to apply to 36.6\% more projects}, \red{thereby} increasing their total earnings in the market.

\begin{table}[h]
\begin{tabular}{ l c c}
\hline
\begin{tabular}[c]{@{}l@{}}Dependent Variable:\\ \textbf{Bid volume}\end{tabular}          & \multicolumn{1}{l}{Coefficient}         & \multicolumn{1}{l}{Standard Error}  \\ \hline
\textit{\final{Avg. first msg. standardization}}                                             & \red{36.6}*** & \red{4.64}                   \\ \hline
\textit{\final{Avg. bid text personalization}}                                         & \red{-8.90 .}  & \red{4.45} \\ \hline
\textit{\final{Avg. bid price (log)}} 
& \red{5.79}*** & \red{.720}  \\ \hline
\textit{Got reviews}                                                              & \red{28.0}*** & \red{2.64}  \\ \hline
\text{Number of observations}                                                   & \multicolumn{2}{c}{161\red{49}}                                                             \\ \hline
\multicolumn{3}{c}{\begin{tabular}[c]{@{}c@{}}*** signifies a p-value \textless .001 and . denotes a p-value \textless .1\\ Errors are clustered by month\end{tabular}} \\ \hline
\end{tabular}
\caption{ Freelancer-level regression predicting bid volume with monthly fixed effects.}
\label{bid_volume}
\end{table}

\section{Discussion}
We examined the effects of standardizing versus personalizing communication practices on individual project success and monthly freelancer earnings. Our first set of findings confirmed that during the bidding stage, content standardization negatively associates with hiring rates \red{(H1a)} while personalization has a positive \red{correlation (H1b)}. From this, we can infer that template use in the initial bidding stage may leave a negative impression with employers by signifying that the associated project is only one among many from the worker's perspective. Relatedly, borrowing and incorporating words and phrases from the client's own description of the project appears to have a favorable effect on clients, perhaps conveying worker sincerity and attentiveness. This suggests that when crafting job proposals (i.e. bid applications), workers may want to carefully read and curate their writing to match the individual job requirements, instead of copying and pasting from templates. Or, as Upwork recommends -- ``Don't use a proposal template" \cite{noauthor_undated-wy}.

However, our analysis of monthly revenue showed that over the long term, content standardization contributes to higher worker earnings, revealing a trade-off between project-level success and long term earning efficiency. To interpret potential mechanisms behind this effect, we examined the effects of the two strategies on bid volume (Table \ref{bid_volume}), which unveiled that using standardized proposal templates enabled workers to submit more bid applications, thereby indirectly contributing to higher monthly earnings. This tradeoff between standardization over the long term and personalization at the individual project level suggests that a worker should keep in mind their broader, long-term career goals while attending to minute and specific details \red{of individual} projects.

Once a freelancer begins working on the project, a fear of losing the gig might cause them to be overly responsive to a particular employer. 
Our results from the execution stage (Table \ref{execution_table}) indicate that this reactive communication approach is negatively associated with project completion \red{(H2b)}. 
This resonates with prior literature on instant messaging, which also found that always being highly responsive to messages in work-related conversations harms workers' abilities to stay on task \cite{Avrahami2004-cd}. Although our analysis does not indicate that response time standardization correlates significantly with long-term earnings, this null result may be due to other hidden characteristics, omitted from our model.

In place of instant replies, freelancers might consider a more proactive form of time-management where they adhere to a consistent daily work schedule and respond only at appropriate times within their own working hours. Naturally, some freelancers may only participate in the online labor market on a part-time basis (referred to by some as \textit{casual earners}), while others are more professionally engaged (including those who are \textit{financially strapped}) \cite{manyika2016independent}. Regardless of a freelancer's online or offline employment status, there is reason to believe that having a consistent work schedule and an increased awareness of time will only benefit a workers' financial and mental well-beings over the long run.

To summarize, our results suggest that freelancers attempting increase their odds of winning a project can consider personalizing the content of their bid applications to cater toward client needs, those who have secured jobs can increase their chances of completion \red{if they refrain from instantly responding to client messages. Workers seeking to increase monthly earnings might consider bidding for projects, which can be achieved through the use of standardized bid proposal templates.} Across all of our models, having reviews on a freelancer's profile positively impacts success, implying that workers seeking to thrive in the online environment may also benefit from image and reputation management.

\subsection{Design Implications}

Given these empirical findings, we propose design recommendations for tools that seek to support gig workers in their various endeavors. Since \red{temporal responsiveness was shown to be harmful toward project completion success,} designers might consider mechanisms that \red{help workers stay focused and on task}. This may take the form of an application or plugin, which may adopt features akin to \red{those found in focus and productivity apps. Current platforms such as Upwork may also want to reconsider the inclusion of responsiveness \footnote{https://support.upwork.com/hc/en-us/articles/211062968-My-Stats} in worker profiles, since a worker's ability to respond to messages quickly might negatively impact their ability to finish a project.}

To make bid personalization easier for workers, tool designers might attempt to use natural-language processing (NLP) methods to extract job requirements from project descriptions and surface them to workers in a more readable fashion. Note that even though current systems do have skill tags that allow clients to clearly define the scope of their project, we can expect many jobs to have unique specifications that cannot be captured by the limited options of a skill tag drop-down. Finally, for workers with relatively low monthly revenues, tools \red{can} provide reminders\ to motivate them to submit more bids and \red{so that they may} maximize their hiring probabilities and work volumes.

\section{Limitations \& Future Work}

Our work is subject to a number of limitations, which also present opportunities for future work. First, our sample is drawn from a single context, a large online freelancing platform. \final{Second, the dataset is from 2010 and analysis of more recent usage of freelancing platforms is needed to determine if patterns of work have since shifted, especially around the increase in remote work due to the 2020 pandemic.} Accordingly, the generalizability of our findings can and should be explored in other settings, such as other online freelancing \red{platforms or more current contexts}. \red{Third}, the observed variation in gig worker strategies that manifest in our data may be subject to some forms of confounding. While our estimations do account for static characteristics of projects, via a set of project fixed effects, future work might explore some of these relationships employing alternative methodologies, e.g., randomized experiments. Finally, our work presently focuses upon communication strategies. However, future work might seek to expand upon this study to consider i) other worker adaptation strategies, e.g., project selection criteria, or ii) the extent to which personalization or standardization may arise and prove effective in other aspects of gig work, beyond communication, e.g., code re-use. 

\section{Conclusion}

We have presented here an examination of gig worker communication strategies on a major online freelancing platform. Using an empirical approach, we suggest cost and benefits of utilizing standardization and personalization techniques at various stages of a project as well as the freelancer's overall career. As the gig economy has continued to grow, particularly with the onset of the COVID-19 pandemic, there is a pressing need to better understand effective work practices, e.g., to inform the design of IT artefacts that may support new gig workers as they seek to situate themselves in their new working environment. As such, it is our hope that future work can expand on this line of inquiry, to offer additional insights that may further aid workers in their pursuit of careers within the gig economy.

\begin{acks}
This work was supported by the National Science Foundation (NSF) under Award 1952085. 
\end{acks}

\bibliographystyle{ACM-Reference-Format}
\bibliography{the-bib}


\begin{thebibliography}{54}


\ifx \showCODEN    \undefined \def \showCODEN     #1{\unskip}     \fi
\ifx \showDOI      \undefined \def \showDOI       #1{#1}\fi
\ifx \showISBNx    \undefined \def \showISBNx     #1{\unskip}     \fi
\ifx \showISBNxiii \undefined \def \showISBNxiii  #1{\unskip}     \fi
\ifx \showISSN     \undefined \def \showISSN      #1{\unskip}     \fi
\ifx \showLCCN     \undefined \def \showLCCN      #1{\unskip}     \fi
\ifx \shownote     \undefined \def \shownote      #1{#1}          \fi
\ifx \showarticletitle \undefined \def \showarticletitle #1{#1}   \fi
\ifx \showURL      \undefined \def \showURL       {\relax}        \fi
\providecommand\bibfield[2]{#2}
\providecommand\bibinfo[2]{#2}
\providecommand\natexlab[1]{#1}
\providecommand\showeprint[2][]{arXiv:#2}

\bibitem[\protect\citeauthoryear{??}{fre}{2021}]%
        {freelance-forward}
 \bibinfo{year}{2021}\natexlab{}.
\newblock \bibinfo{title}{Freelance Forward 2020}.
\newblock
  \bibinfo{howpublished}{\url{https://www.upwork.com/i/freelance-forward}}.
\newblock
\newblock
\shownote{Accessed: 2021-9-1.}


\bibitem[\protect\citeauthoryear{??}{noa}{2021}]%
        {noauthor_undated-wy}
 \bibinfo{year}{2021}\natexlab{}.
\newblock \bibinfo{title}{How to create a proposal that wins jobs}.
\newblock
  \bibinfo{howpublished}{\url{https://www.upwork.com/resources/how-to-create-a-proposal-that-wins-jobs}}.
\newblock
\newblock
\shownote{Accessed: 2021-9-1.}


\bibitem[\protect\citeauthoryear{Ashford, Caza, and Reid}{Ashford
  et~al\mbox{.}}{2018}]%
        {Ashford2018-dw}
\bibfield{author}{\bibinfo{person}{Susan~J Ashford},
  \bibinfo{person}{Brianna~Barker Caza}, {and} \bibinfo{person}{Erin~M Reid}.}
  \bibinfo{year}{2018}\natexlab{}.
\newblock \showarticletitle{{From surviving to thriving in the gig economy: A
  research agenda for individuals in the new world of work}}.
\newblock \bibinfo{journal}{\emph{Research in Organizational Behavior}}
  \bibinfo{volume}{38} (\bibinfo{year}{2018}), \bibinfo{pages}{23--41}.
\newblock


\bibitem[\protect\citeauthoryear{Atarah, Peprah, Amartey, and Bamfo}{Atarah
  et~al\mbox{.}}{2021}]%
        {Atarah2021-ox}
\bibfield{author}{\bibinfo{person}{Bede~Akorige Atarah},
  \bibinfo{person}{Augustine~Awuah Peprah}, \bibinfo{person}{Abednego F~Okoe
  Amartey}, {and} \bibinfo{person}{Bylon~Abeeku Bamfo}.}
  \bibinfo{year}{2021}\natexlab{}.
\newblock \showarticletitle{{Making do by doing without: bricolage in the
  funding sources of female entrepreneurs in resource-constrained
  environments}}.
\newblock \bibinfo{journal}{\emph{Journal of Global Entrepreneurship Research}}
  (\bibinfo{year}{2021}), \bibinfo{pages}{1--18}.
\newblock


\bibitem[\protect\citeauthoryear{Avrahami and Hudson}{Avrahami and
  Hudson}{2004}]%
        {Avrahami2004-cd}
\bibfield{author}{\bibinfo{person}{Daniel Avrahami} {and}
  \bibinfo{person}{Scott~E Hudson}.} \bibinfo{year}{2004}\natexlab{}.
\newblock \showarticletitle{{{QnA}: augmenting an instant messaging client to
  balance user responsiveness and performance}}.
\newblock \bibinfo{journal}{\emph{Proceedings of the 2004 ACM conference on
  Computer supported cooperative work - CSCW '04}} (\bibinfo{year}{2004}),
  \bibinfo{pages}{515--518}.
\newblock


\bibitem[\protect\citeauthoryear{Barrero, Bloom, and Davis}{Barrero
  et~al\mbox{.}}{2020}]%
        {Barrero2020-sf}
\bibfield{author}{\bibinfo{person}{Jose~Maria Barrero},
  \bibinfo{person}{Nicholas Bloom}, {and} \bibinfo{person}{Steven~J Davis}.}
  \bibinfo{year}{2020}\natexlab{}.
\newblock \showarticletitle{{Why Working From Home Will Stick}}.
\newblock \bibinfo{journal}{\emph{SSRN Electronic Journal}}
  (\bibinfo{year}{2020}).
\newblock


\bibitem[\protect\citeauthoryear{Blaising, Kotturi, Kulkarni, and
  Dabbish}{Blaising et~al\mbox{.}}{2021}]%
        {Blaising2021-bn}
\bibfield{author}{\bibinfo{person}{Allie Blaising}, \bibinfo{person}{Yasmine
  Kotturi}, \bibinfo{person}{Chinmay Kulkarni}, {and} \bibinfo{person}{Laura
  Dabbish}.} \bibinfo{year}{2021}\natexlab{}.
\newblock \showarticletitle{{Making it Work, or Not}}.
\newblock \bibinfo{journal}{\emph{Proceedings of the ACM on Human-Computer
  Interaction}} \bibinfo{volume}{4}, \bibinfo{number}{CSCW3}
  (\bibinfo{year}{2021}), \bibinfo{pages}{1--29}.
\newblock


\bibitem[\protect\citeauthoryear{Blom}{Blom}{2000}]%
        {Blom_2000}
\bibfield{author}{\bibinfo{person}{Jan Blom}.} \bibinfo{year}{2000}\natexlab{}.
\newblock \showarticletitle{Personalization: a taxonomy}.
\newblock \bibinfo{journal}{\emph{CHI ’00 extended abstracts on Human factors
  in computing systems - CHI ’00}} (\bibinfo{year}{2000}),
  \bibinfo{pages}{313}.
\newblock
\urldef\tempurl%
\url{https://doi.org/10.1145/633292.633483}
\showDOI{\tempurl}


\bibitem[\protect\citeauthoryear{Brewster, Fitzpatrick, Cox, Kostakos, Lascau,
  Gould, Cox, Karmannaya, and Brumby}{Brewster et~al\mbox{.}}{2019}]%
  {Brewster_Fitzpatrick_Cox_Kostakos_Lascau_Gould_Cox_Karmannaya_Brumby_2019}
\bibfield{author}{\bibinfo{person}{Stephen Brewster},
  \bibinfo{person}{Geraldine Fitzpatrick}, \bibinfo{person}{Anna Cox},
  \bibinfo{person}{Vassilis Kostakos}, \bibinfo{person}{Laura Lascau},
  \bibinfo{person}{Sandy J~J Gould}, \bibinfo{person}{Anna~L Cox},
  \bibinfo{person}{Elizaveta Karmannaya}, {and} \bibinfo{person}{Duncan~P
  Brumby}.} \bibinfo{year}{2019}\natexlab{}.
\newblock \showarticletitle{Monotasking or Multitasking: Designing for
  Crowdworkers’ Preferences}.
\newblock \bibinfo{journal}{\emph{Proceedings of the 2019 CHI Conference on
  Human Factors in Computing Systems}} (\bibinfo{year}{2019}),
  \bibinfo{pages}{1–14}.
\newblock
\urldef\tempurl%
\url{https://doi.org/10.1145/3290605.3300649}
\showDOI{\tempurl}


\bibitem[\protect\citeauthoryear{Busch and Barkema}{Busch and Barkema}{2021}]%
        {Busch2021-ls}
\bibfield{author}{\bibinfo{person}{Christian Busch} {and}
  \bibinfo{person}{Harry Barkema}.} \bibinfo{year}{2021}\natexlab{}.
\newblock \showarticletitle{{From necessity to opportunity: Scaling bricolage
  across resource constrained environments}}.
\newblock \bibinfo{journal}{\emph{Strategic Manage. J.}} \bibinfo{volume}{42},
  \bibinfo{number}{4} (\bibinfo{year}{2021}), \bibinfo{pages}{741--773}.
\newblock


\bibitem[\protect\citeauthoryear{Chang and Webster}{Chang and Webster}{2019}]%
        {Chang2019-ri}
\bibfield{author}{\bibinfo{person}{Frances Chang} {and}
  \bibinfo{person}{Cynthia~M Webster}.} \bibinfo{year}{2019}\natexlab{}.
\newblock \showarticletitle{{Effects of Network Bricolage on Entrepreneurs'
  Resource Creation}}.
\newblock \bibinfo{journal}{\emph{Academy of Management Proceedings}}
  \bibinfo{volume}{2019}, \bibinfo{number}{1} (\bibinfo{year}{2019}),
  \bibinfo{pages}{10137}.
\newblock


\bibitem[\protect\citeauthoryear{Churchill}{Churchill}{2013}]%
        {Churchill_2013}
\bibfield{author}{\bibinfo{person}{Elizabeth~F. Churchill}.}
  \bibinfo{year}{2013}\natexlab{}.
\newblock \showarticletitle{Putting the person back into personalization}.
\newblock \bibinfo{journal}{\emph{interactions}} \bibinfo{volume}{20},
  \bibinfo{number}{5} (\bibinfo{year}{2013}), \bibinfo{pages}{12–15}.
\newblock
\showISSN{1072-5520}
\urldef\tempurl%
\url{https://doi.org/10.1145/2504847}
\showDOI{\tempurl}


\bibitem[\protect\citeauthoryear{Cutrell, Czerwinski, and Horvitz}{Cutrell
  et~al\mbox{.}}{2000}]%
        {Cutrell2000-ap}
\bibfield{author}{\bibinfo{person}{Edward~B Cutrell}, \bibinfo{person}{Mary
  Czerwinski}, {and} \bibinfo{person}{Eric Horvitz}.}
  \bibinfo{year}{2000}\natexlab{}.
\newblock \showarticletitle{{Effects of instant messaging interruptions on
  computing tasks}}.
\newblock \bibinfo{journal}{\emph{CHI '00 extended abstracts on Human factors
  in computing systems - CHI '00}} (\bibinfo{year}{2000}),
  \bibinfo{pages}{99--100}.
\newblock


\bibitem[\protect\citeauthoryear{Dillahunt, Wang, Wheeler, Cheng, Hecht, and
  Zhu}{Dillahunt et~al\mbox{.}}{2017}]%
        {Dillahunt_Wang_Wheeler_Cheng_Hecht_Zhu_2017}
\bibfield{author}{\bibinfo{person}{Tawanna~R Dillahunt}, \bibinfo{person}{Xinyi
  Wang}, \bibinfo{person}{Earnest Wheeler}, \bibinfo{person}{Hao~Fei Cheng},
  \bibinfo{person}{Brent Hecht}, {and} \bibinfo{person}{Haiyi Zhu}.}
  \bibinfo{year}{2017}\natexlab{}.
\newblock \showarticletitle{The Sharing Economy in Computing: A Systematic
  Literature Review}.
\newblock \bibinfo{journal}{\emph{Proceedings of the ACM on Human-Computer
  Interaction}} \bibinfo{volume}{1}, \bibinfo{number}{CSCW}
  (\bibinfo{year}{2017}), \bibinfo{pages}{1–26}.
\newblock
\urldef\tempurl%
\url{https://doi.org/10.1145/3134673}
\showDOI{\tempurl}


\bibitem[\protect\citeauthoryear{Dunn}{Dunn}{2017}]%
        {dunn2017digital}
\bibfield{author}{\bibinfo{person}{Michael Dunn}.}
  \bibinfo{year}{2017}\natexlab{}.
\newblock \showarticletitle{Digital work: New opportunities or lost wages?}
\newblock \bibinfo{journal}{\emph{American Journal of Management}}
  \bibinfo{volume}{17}, \bibinfo{number}{4} (\bibinfo{year}{2017}),
  \bibinfo{pages}{10--27}.
\newblock


\bibitem[\protect\citeauthoryear{Dunn, Stephany, Sawyer, Munoz, Raheja,
  Vaccaro, and Lehdonvirta}{Dunn et~al\mbox{.}}{2020}]%
        {dunn2020motivation}
\bibfield{author}{\bibinfo{person}{Michael Dunn}, \bibinfo{person}{Fabian
  Stephany}, \bibinfo{person}{Steven Sawyer}, \bibinfo{person}{Isabel Munoz},
  \bibinfo{person}{Raghav Raheja}, \bibinfo{person}{Gabrielle Vaccaro}, {and}
  \bibinfo{person}{Vili Lehdonvirta}.} \bibinfo{year}{2020}\natexlab{}.
\newblock \showarticletitle{When motivation becomes desperation: Online
  freelancing during the Covid-19 pandemic}.
\newblock  (\bibinfo{year}{2020}).
\newblock


\bibitem[\protect\citeauthoryear{Dyche, O'Brien, et~al\mbox{.}}{Dyche
  et~al\mbox{.}}{2002}]%
        {dyche2002crm}
\bibfield{author}{\bibinfo{person}{Jill Dyche}, \bibinfo{person}{Mary~Mary
  O'Brien}, {et~al\mbox{.}}} \bibinfo{year}{2002}\natexlab{}.
\newblock \bibinfo{booktitle}{\emph{The CRM handbook: A business guide to
  customer relationship management}}.
\newblock \bibinfo{publisher}{Addison-Wesley Professional}.
\newblock


\bibitem[\protect\citeauthoryear{Evers, Naaman, Fitzpatrick, Karahalios,
  Lampinen, Monroy-Hernández, Lampinen, Lutz, Newlands, Light, and
  et~al.}{Evers et~al\mbox{.}}{2018}]%
        {power_struggles}
\bibfield{author}{\bibinfo{person}{Vanessa Evers}, \bibinfo{person}{Mor
  Naaman}, \bibinfo{person}{Geraldine Fitzpatrick}, \bibinfo{person}{Karrie
  Karahalios}, \bibinfo{person}{Airi Lampinen}, \bibinfo{person}{Andrés
  Monroy-Hernández}, \bibinfo{person}{Airi Lampinen},
  \bibinfo{person}{Christoph Lutz}, \bibinfo{person}{Gemma Newlands},
  \bibinfo{person}{Ann Light}, {and} \bibinfo{person}{et al.}}
  \bibinfo{year}{2018}\natexlab{}.
\newblock \showarticletitle{Power Struggles in the Digital Economy}.
\newblock \bibinfo{journal}{\emph{Companion of the 2018 ACM Conference on
  Computer Supported Cooperative Work and Social Computing}}
  \bibinfo{volume}{29}, \bibinfo{number}{1–2} (\bibinfo{year}{2018}),
  \bibinfo{pages}{417–423}.
\newblock
\showISSN{0925-9724}
\urldef\tempurl%
\url{https://doi.org/10.1145/3272973.3273004}
\showDOI{\tempurl}


\bibitem[\protect\citeauthoryear{Fan and Poole}{Fan and Poole}{2006}]%
        {fan2006personalization}
\bibfield{author}{\bibinfo{person}{Haiyan Fan} {and}
  \bibinfo{person}{Marshall~Scott Poole}.} \bibinfo{year}{2006}\natexlab{}.
\newblock \showarticletitle{What is personalization? Perspectives on the design
  and implementation of personalization in information systems}.
\newblock \bibinfo{journal}{\emph{Journal of Organizational Computing and
  Electronic Commerce}} \bibinfo{volume}{16}, \bibinfo{number}{3-4}
  (\bibinfo{year}{2006}), \bibinfo{pages}{179--202}.
\newblock


\bibitem[\protect\citeauthoryear{Griffith, Hu, and Ryans}{Griffith
  et~al\mbox{.}}{2000}]%
        {Griffith}
\bibfield{author}{\bibinfo{person}{David~A. Griffith},
  \bibinfo{person}{Michael~Y. Hu}, {and} \bibinfo{person}{John~K. Ryans}.}
  \bibinfo{year}{2000}\natexlab{}.
\newblock \showarticletitle{Process Standardization across Intra- and
  Inter-Cultural Relationships}.
\newblock \bibinfo{journal}{\emph{Journal of International Business Studies}}
  \bibinfo{volume}{31}, \bibinfo{number}{2} (\bibinfo{year}{2000}),
  \bibinfo{pages}{303–324}.
\newblock
\showISSN{0047-2506}
\urldef\tempurl%
\url{https://doi.org/10.1057/palgrave.jibs.8490908}
\showDOI{\tempurl}


\bibitem[\protect\citeauthoryear{Hall and Krueger}{Hall and Krueger}{2018}]%
        {Hall2018-uo}
\bibfield{author}{\bibinfo{person}{Jonathan~V Hall} {and}
  \bibinfo{person}{Alan~B Krueger}.} \bibinfo{year}{2018}\natexlab{}.
\newblock \showarticletitle{{An Analysis of the Labor Market for Uber's
  {Driver-Partners} in the United States}}.
\newblock \bibinfo{journal}{\emph{ILR Review}} \bibinfo{volume}{71},
  \bibinfo{number}{3} (\bibinfo{year}{2018}), \bibinfo{pages}{705--732}.
\newblock


\bibitem[\protect\citeauthoryear{Hanseth, Monteiro, and Hatling}{Hanseth
  et~al\mbox{.}}{1996}]%
        {Hanseth1996-ux}
\bibfield{author}{\bibinfo{person}{Ole Hanseth}, \bibinfo{person}{Eric
  Monteiro}, {and} \bibinfo{person}{Morten Hatling}.}
  \bibinfo{year}{1996}\natexlab{}.
\newblock \showarticletitle{Developing Information Infrastructure: The Tension
  Between Standardization and Flexibility}.
\newblock \bibinfo{journal}{\emph{Sci. Technol. Human Values}}
  \bibinfo{volume}{21}, \bibinfo{number}{4} (\bibinfo{date}{Oct.}
  \bibinfo{year}{1996}), \bibinfo{pages}{407--426}.
\newblock


\bibitem[\protect\citeauthoryear{Hanson}{Hanson}{2000}]%
        {hanson2000principles}
\bibfield{author}{\bibinfo{person}{Ward Hanson}.}
  \bibinfo{year}{2000}\natexlab{}.
\newblock \showarticletitle{Principles of Internet Marketing Ward Hanson}.
\newblock  (\bibinfo{year}{2000}).
\newblock


\bibitem[\protect\citeauthoryear{Hong, Peng, Burtch, and Huang}{Hong
  et~al\mbox{.}}{2021}]%
        {Hong2021-hj}
\bibfield{author}{\bibinfo{person}{Yili Hong}, \bibinfo{person}{Jing Peng},
  \bibinfo{person}{Gordon Burtch}, {and} \bibinfo{person}{Ni Huang}.}
  \bibinfo{year}{2021}\natexlab{}.
\newblock \showarticletitle{{Just {DM} Me (Politely): Direct Messaging,
  Politeness, and Hiring Outcomes in Online Labor Markets}}.
\newblock \bibinfo{journal}{\emph{Information Systems Research}}
  (\bibinfo{year}{2021}).
\newblock


\bibitem[\protect\citeauthoryear{Horton}{Horton}{2018}]%
        {Horton2018-eq}
\bibfield{author}{\bibinfo{person}{John~J Horton}.}
  \bibinfo{year}{2018}\natexlab{}.
\newblock \showarticletitle{{Buyer Uncertainty About Seller Capacity: Causes,
  Consequences, and a Partial Solution}}.
\newblock \bibinfo{journal}{\emph{SSRN Electronic Journal}}
  (\bibinfo{year}{2018}).
\newblock


\bibitem[\protect\citeauthoryear{Huang, Burtch, Hong, and Pavlou}{Huang
  et~al\mbox{.}}{2020}]%
        {huang2020unemployment}
\bibfield{author}{\bibinfo{person}{Ni Huang}, \bibinfo{person}{Gordon Burtch},
  \bibinfo{person}{Yili Hong}, {and} \bibinfo{person}{Paul~A Pavlou}.}
  \bibinfo{year}{2020}\natexlab{}.
\newblock \showarticletitle{Unemployment and worker participation in the gig
  economy: Evidence from an online labor market}.
\newblock \bibinfo{journal}{\emph{Information Systems Research}}
  \bibinfo{volume}{31}, \bibinfo{number}{2} (\bibinfo{year}{2020}),
  \bibinfo{pages}{431--448}.
\newblock


\bibitem[\protect\citeauthoryear{Jarrahi and Sutherland}{Jarrahi and
  Sutherland}{2019}]%
        {Jarrahi_Sutherland_2019}
\bibfield{author}{\bibinfo{person}{Mohammad~Hossein Jarrahi} {and}
  \bibinfo{person}{Will Sutherland}.} \bibinfo{year}{2019}\natexlab{}.
\newblock \showarticletitle{Information in Contemporary Society, 14th
  International Conference, iConference 2019, Washington, DC, USA, March
  31–April 3, 2019, Proceedings}.
\newblock \bibinfo{journal}{\emph{Lecture Notes in Computer Science}}
  (\bibinfo{year}{2019}), \bibinfo{pages}{578–589}.
\newblock
\showISSN{0302-9743}
\urldef\tempurl%
\url{https://doi.org/10.1007/978-3-030-15742-5_55}
\showDOI{\tempurl}


\bibitem[\protect\citeauthoryear{Jarrahi, Sutherland, Nelson, and
  Sawyer}{Jarrahi et~al\mbox{.}}{2020}]%
        {platform_manage}
\bibfield{author}{\bibinfo{person}{Mohammad~Hossein Jarrahi},
  \bibinfo{person}{Will Sutherland}, \bibinfo{person}{Sarah~Beth Nelson}, {and}
  \bibinfo{person}{Steve Sawyer}.} \bibinfo{year}{2020}\natexlab{}.
\newblock \showarticletitle{Platformic Management, Boundary Resources for Gig
  Work, and Worker Autonomy}.
\newblock \bibinfo{journal}{\emph{Computer Supported Cooperative Work (CSCW)}}
  \bibinfo{volume}{29}, \bibinfo{number}{1–2} (\bibinfo{year}{2020}),
  \bibinfo{pages}{153–189}.
\newblock
\showISSN{0925-9724}
\urldef\tempurl%
\url{https://doi.org/10.1007/s10606-019-09368-7}
\showDOI{\tempurl}


\bibitem[\protect\citeauthoryear{Kaptein, Markopoulos, Ruyter, and
  Aarts}{Kaptein et~al\mbox{.}}{2015}]%
        {Kaptein_Markopoulos_Ruyter_Aarts_2015}
\bibfield{author}{\bibinfo{person}{Maurits Kaptein}, \bibinfo{person}{Panos
  Markopoulos}, \bibinfo{person}{Boris~de Ruyter}, {and} \bibinfo{person}{Emile
  Aarts}.} \bibinfo{year}{2015}\natexlab{}.
\newblock \showarticletitle{Personalizing persuasive technologies: Explicit and
  implicit personalization using persuasion profiles}.
\newblock \bibinfo{journal}{\emph{International Journal of Human-Computer
  Studies}}  \bibinfo{volume}{77} (\bibinfo{year}{2015}),
  \bibinfo{pages}{38–51}.
\newblock
\showISSN{1071-5819}
\urldef\tempurl%
\url{https://doi.org/10.1016/j.ijhcs.2015.01.004}
\showDOI{\tempurl}


\bibitem[\protect\citeauthoryear{Kingsley, Gray, and Suri}{Kingsley
  et~al\mbox{.}}{2015}]%
        {kingsley2015accounting}
\bibfield{author}{\bibinfo{person}{Sara~Constance Kingsley},
  \bibinfo{person}{Mary~L Gray}, {and} \bibinfo{person}{Siddharth Suri}.}
  \bibinfo{year}{2015}\natexlab{}.
\newblock \showarticletitle{Accounting for market frictions and power
  asymmetries in online labor markets}.
\newblock \bibinfo{journal}{\emph{Policy \& Internet}} \bibinfo{volume}{7},
  \bibinfo{number}{4} (\bibinfo{year}{2015}), \bibinfo{pages}{383--400}.
\newblock


\bibitem[\protect\citeauthoryear{Kost, Fieseler, and Wong}{Kost
  et~al\mbox{.}}{2020}]%
        {boundary}
\bibfield{author}{\bibinfo{person}{Dominique Kost}, \bibinfo{person}{Christian
  Fieseler}, {and} \bibinfo{person}{Sut~I Wong}.}
  \bibinfo{year}{2020}\natexlab{}.
\newblock \showarticletitle{{Boundaryless careers in the gig economy: An
  oxymoron?}}
\newblock \bibinfo{journal}{\emph{Human Resource Management Journal}}
  \bibinfo{volume}{30}, \bibinfo{number}{1} (\bibinfo{year}{2020}),
  \bibinfo{pages}{100--113}.
\newblock


\bibitem[\protect\citeauthoryear{Kuhn and Galloway}{Kuhn and Galloway}{2019}]%
        {Kuhn2019-fi}
\bibfield{author}{\bibinfo{person}{Kristine~M Kuhn} {and}
  \bibinfo{person}{Tera~L Galloway}.} \bibinfo{year}{2019}\natexlab{}.
\newblock \showarticletitle{{Expanding perspectives on gig work and gig
  workers}}.
\newblock \bibinfo{journal}{\emph{Journal of Managerial Psychology}}
  \bibinfo{volume}{34}, \bibinfo{number}{4} (\bibinfo{year}{2019}),
  \bibinfo{pages}{186--191}.
\newblock


\bibitem[\protect\citeauthoryear{Lavin and Maynard}{Lavin and Maynard}{2001}]%
        {rapport}
\bibfield{author}{\bibinfo{person}{Danielle Lavin} {and}
  \bibinfo{person}{Douglas~W Maynard}.} \bibinfo{year}{2001}\natexlab{}.
\newblock \showarticletitle{Standardization vs. Rapport: Respondent Laughter
  and Interviewer Reaction during Telephone Surveys}.
\newblock \bibinfo{journal}{\emph{American Sociological Review}}
  \bibinfo{volume}{66}, \bibinfo{number}{3} (\bibinfo{year}{2001}),
  \bibinfo{pages}{453}.
\newblock
\showISSN{0003-1224}
\urldef\tempurl%
\url{https://doi.org/10.2307/3088888}
\showDOI{\tempurl}


\bibitem[\protect\citeauthoryear{Li and Karahanna}{Li and Karahanna}{2015}]%
        {consortium}
\bibfield{author}{\bibinfo{person}{Seth~Siyuan Li} {and} \bibinfo{person}{Elena
  Karahanna}.} \bibinfo{year}{2015}\natexlab{}.
\newblock \showarticletitle{Online recommendation systems in a B2C E-commerce
  context: a review and future directions}.
\newblock \bibinfo{journal}{\emph{Journal of the Association for Information
  Systems}} \bibinfo{volume}{16}, \bibinfo{number}{2} (\bibinfo{year}{2015}),
  \bibinfo{pages}{2}.
\newblock


\bibitem[\protect\citeauthoryear{Lim, Sia, Lee, and Benbasat}{Lim
  et~al\mbox{.}}{2006}]%
        {trust}
\bibfield{author}{\bibinfo{person}{Kai~H Lim}, \bibinfo{person}{Choon~Ling
  Sia}, \bibinfo{person}{Matthew~KO Lee}, {and} \bibinfo{person}{Izak
  Benbasat}.} \bibinfo{year}{2006}\natexlab{}.
\newblock \showarticletitle{Do I trust you online, and if so, will I buy? An
  empirical study of two trust-building strategies}.
\newblock \bibinfo{journal}{\emph{Journal of management information systems}}
  \bibinfo{volume}{23}, \bibinfo{number}{2} (\bibinfo{year}{2006}),
  \bibinfo{pages}{233--266}.
\newblock


\bibitem[\protect\citeauthoryear{Lvi-Strauss}{Lvi-Strauss}{1966}]%
        {lvi1966savage}
\bibfield{author}{\bibinfo{person}{Claude Lvi-Strauss}.}
  \bibinfo{year}{1966}\natexlab{}.
\newblock \bibinfo{booktitle}{\emph{The savage mind}}.
\newblock \bibinfo{publisher}{University of Chicago Press}.
\newblock


\bibitem[\protect\citeauthoryear{Manyika, Lund, Bughin, Robinson, Mischke, and
  Mahajan}{Manyika et~al\mbox{.}}{2016}]%
        {manyika2016independent}
\bibfield{author}{\bibinfo{person}{James Manyika}, \bibinfo{person}{Susan
  Lund}, \bibinfo{person}{Jacques Bughin}, \bibinfo{person}{Kelsey Robinson},
  \bibinfo{person}{Jan Mischke}, {and} \bibinfo{person}{Deepa Mahajan}.}
  \bibinfo{year}{2016}\natexlab{}.
\newblock
  \bibinfo{booktitle}{\emph{Independent-Work-Choice-necessity-and-the-gig-economy}}.
\newblock \bibinfo{type}{{T}echnical {R}eport}. \bibinfo{institution}{McKinsey
  Global Institute}.
\newblock


\bibitem[\protect\citeauthoryear{M{\"u}nstermann, Eckhardt, and
  Weitzel}{M{\"u}nstermann et~al\mbox{.}}{2010}]%
        {Munstermann2010-yf}
\bibfield{author}{\bibinfo{person}{Bj{\"o}rn M{\"u}nstermann},
  \bibinfo{person}{Andreas Eckhardt}, {and} \bibinfo{person}{Tim Weitzel}.}
  \bibinfo{year}{2010}\natexlab{}.
\newblock \bibinfo{title}{The performance impact of business process
  standardization}.
\newblock , \bibinfo{numpages}{29--56}~pages.
\newblock


\bibitem[\protect\citeauthoryear{M{\"u}nstermann and Weitzel}{M{\"u}nstermann
  and Weitzel}{2008}]%
        {Munstermann2008-yj}
\bibfield{author}{\bibinfo{person}{Bj{\"o}rn M{\"u}nstermann} {and}
  \bibinfo{person}{Tim Weitzel}.} \bibinfo{year}{2008}\natexlab{}.
\newblock \showarticletitle{What is process standardization?}. In
  \bibinfo{booktitle}{\emph{{CONF-IRM} 2008 Proceedings}}. \bibinfo{pages}{64}.
\newblock


\bibitem[\protect\citeauthoryear{Newlands, Lutz, and Fieseler}{Newlands
  et~al\mbox{.}}{2017}]%
        {power}
\bibfield{author}{\bibinfo{person}{Gemma Newlands}, \bibinfo{person}{Christoph
  Lutz}, {and} \bibinfo{person}{Christian Fieseler}.}
  \bibinfo{year}{2017}\natexlab{}.
\newblock \showarticletitle{Power in the Sharing Economy: European
  Perspectives}.
\newblock \bibinfo{journal}{\emph{SSRN Electronic Journal}}
  (\bibinfo{year}{2017}).
\newblock
\urldef\tempurl%
\url{https://doi.org/10.2139/ssrn.3046473}
\showDOI{\tempurl}


\bibitem[\protect\citeauthoryear{Newman}{Newman}{2016}]%
        {Newman2016-nd}
\bibfield{author}{\bibinfo{person}{Amy Newman}.}
  \bibinfo{year}{2016}\natexlab{}.
\newblock \showarticletitle{Communication Planning: A Template for
  Organizational Change}.
\newblock  (\bibinfo{date}{Feb.} \bibinfo{year}{2016}).
\newblock


\bibitem[\protect\citeauthoryear{Peppers and Rogers}{Peppers and
  Rogers}{1997}]%
        {peppers}
\bibfield{author}{\bibinfo{person}{Don Peppers} {and} \bibinfo{person}{Martha
  Rogers}.} \bibinfo{year}{1997}\natexlab{}.
\newblock \bibinfo{booktitle}{\emph{Enterprise one to one: Tools for competing
  in the interactive age}}.
\newblock \bibinfo{publisher}{Currency Doubleday New York}.
\newblock


\bibitem[\protect\citeauthoryear{Ramakumar and Cooper}{Ramakumar and
  Cooper}{2004}]%
        {Ramakumar2004-fv}
\bibfield{author}{\bibinfo{person}{Arun Ramakumar} {and}
  \bibinfo{person}{Blaine Cooper}.} \bibinfo{year}{2004}\natexlab{}.
\newblock \showarticletitle{Process Standardization Proves Profitable}.
\newblock \bibinfo{journal}{\emph{Quality; Troy}} \bibinfo{volume}{43},
  \bibinfo{number}{2} (\bibinfo{date}{Feb.} \bibinfo{year}{2004}),
  \bibinfo{pages}{42--45}.
\newblock


\bibitem[\protect\citeauthoryear{Roberts and Zahay}{Roberts and Zahay}{2012}]%
        {roberts2012internet}
\bibfield{author}{\bibinfo{person}{Mary~Lou Roberts} {and}
  \bibinfo{person}{Debra Zahay}.} \bibinfo{year}{2012}\natexlab{}.
\newblock \bibinfo{booktitle}{\emph{Internet marketing: Integrating online and
  offline strategies}}.
\newblock \bibinfo{publisher}{Cengage Learning}.
\newblock


\bibitem[\protect\citeauthoryear{Rosenblat and Stark}{Rosenblat and
  Stark}{2016}]%
        {Rosenblat_Stark_2016}
\bibfield{author}{\bibinfo{person}{Alex Rosenblat} {and} \bibinfo{person}{Luke
  Stark}.} \bibinfo{year}{2016}\natexlab{}.
\newblock \showarticletitle{Algorithmic Labor and Information Asymmetries: A
  Case Study of Uber’s Drivers}.
\newblock \bibinfo{journal}{\emph{SSRN Electronic Journal}}
  (\bibinfo{year}{2016}).
\newblock
\urldef\tempurl%
\url{https://doi.org/10.2139/ssrn.2686227}
\showDOI{\tempurl}


\bibitem[\protect\citeauthoryear{Shevchuk, Strebkov, and Davis}{Shevchuk
  et~al\mbox{.}}{2019}]%
        {Shevchuk_Strebkov_Davis_2019}
\bibfield{author}{\bibinfo{person}{Andrey Shevchuk}, \bibinfo{person}{Denis
  Strebkov}, {and} \bibinfo{person}{Shannon~N. Davis}.}
  \bibinfo{year}{2019}\natexlab{}.
\newblock \showarticletitle{The Autonomy Paradox: How Night Work Undermines
  Subjective Well-Being of Internet-Based Freelancers}.
\newblock \bibinfo{journal}{\emph{ILR Review}} \bibinfo{volume}{72},
  \bibinfo{number}{1} (\bibinfo{year}{2019}), \bibinfo{pages}{75–100}.
\newblock
\showISSN{0019-7939}
\urldef\tempurl%
\url{https://doi.org/10.1177/0019793918767114}
\showDOI{\tempurl}


\bibitem[\protect\citeauthoryear{Siqueira and Herring}{Siqueira and
  Herring}{2009}]%
        {Siqueira2009-md}
\bibfield{author}{\bibinfo{person}{Amaury~de Siqueira} {and}
  \bibinfo{person}{Susan~C Herring}.} \bibinfo{year}{2009}\natexlab{}.
\newblock \showarticletitle{{Temporal Patterns in {Student-Advisor} Instant
  Messaging Exchanges: Individual Variation and Accommodation}}.
\newblock \bibinfo{journal}{\emph{2009 42nd Hawaii International Conference on
  System Sciences}}  \bibinfo{volume}{1} (\bibinfo{year}{2009}),
  \bibinfo{pages}{1--10}.
\newblock


\bibitem[\protect\citeauthoryear{Surprenant and Solomon}{Surprenant and
  Solomon}{1987}]%
        {Surprenant1987-sz}
\bibfield{author}{\bibinfo{person}{Carol~F Surprenant} {and}
  \bibinfo{person}{Michael~R Solomon}.} \bibinfo{year}{1987}\natexlab{}.
\newblock \showarticletitle{{Predictability and Personalization in the Service
  Encounter}}.
\newblock \bibinfo{journal}{\emph{J. Mark.}} \bibinfo{volume}{51},
  \bibinfo{number}{2} (\bibinfo{year}{1987}), \bibinfo{pages}{86--96}.
\newblock


\bibitem[\protect\citeauthoryear{Sutherland, Jarrahi, Dunn, and
  Nelson}{Sutherland et~al\mbox{.}}{2020}]%
        {Sutherland2020-wk}
\bibfield{author}{\bibinfo{person}{Will Sutherland},
  \bibinfo{person}{Mohammad~Hossein Jarrahi}, \bibinfo{person}{Michael Dunn},
  {and} \bibinfo{person}{Sarah~Beth Nelson}.} \bibinfo{year}{2020}\natexlab{}.
\newblock \showarticletitle{{Work Precarity and Gig Literacies in Online
  Freelancing}}.
\newblock \bibinfo{journal}{\emph{Work Employ. Soc.}} \bibinfo{volume}{34},
  \bibinfo{number}{3} (\bibinfo{year}{2020}), \bibinfo{pages}{457--475}.
\newblock


\bibitem[\protect\citeauthoryear{Tan, Wang, and Tan}{Tan et~al\mbox{.}}{2019}]%
        {alibaba}
\bibfield{author}{\bibinfo{person}{Xue Tan}, \bibinfo{person}{Youwei Wang},
  {and} \bibinfo{person}{Yong Tan}.} \bibinfo{year}{2019}\natexlab{}.
\newblock \showarticletitle{{Impact of Live Chat on Purchase in Electronic
  Markets: The Moderating Role of Information Cues}}.
\newblock \bibinfo{journal}{\emph{SSRN Electronic Journal}}
  (\bibinfo{year}{2019}).
\newblock


\bibitem[\protect\citeauthoryear{Vanevenhoven, Winkel, Malewicki, Dougan, and
  Bronson}{Vanevenhoven et~al\mbox{.}}{2011}]%
        {Vanevenhoven2011-bx}
\bibfield{author}{\bibinfo{person}{Jeff Vanevenhoven}, \bibinfo{person}{Doan
  Winkel}, \bibinfo{person}{Debra Malewicki}, \bibinfo{person}{William~L
  Dougan}, {and} \bibinfo{person}{James Bronson}.}
  \bibinfo{year}{2011}\natexlab{}.
\newblock \showarticletitle{{Varieties of bricolage and the process of
  entrepreneurship}}.
\newblock \bibinfo{journal}{\emph{New England Journal of Entrepreneurship}}
  \bibinfo{volume}{14}, \bibinfo{number}{2} (\bibinfo{year}{2011}),
  \bibinfo{pages}{53--66}.
\newblock


\bibitem[\protect\citeauthoryear{Wuellenweber, Koenig, Beimborn, and
  Weitzel}{Wuellenweber et~al\mbox{.}}{2009}]%
        {Wuellenweber2009-pa}
\bibfield{author}{\bibinfo{person}{Kim Wuellenweber}, \bibinfo{person}{Wolfgang
  Koenig}, \bibinfo{person}{Daniel Beimborn}, {and} \bibinfo{person}{Tim
  Weitzel}.} \bibinfo{year}{2009}\natexlab{}.
\newblock \showarticletitle{The Impact of Process Standardization on Business
  Process Outsourcing Success}.
\newblock In \bibinfo{booktitle}{\emph{Information Systems Outsourcing:
  Enduring Themes, Global Challenges, and Process Opportunities}},
  \bibfield{editor}{\bibinfo{person}{Rudy Hirschheim}, \bibinfo{person}{Armin
  Heinzl}, {and} \bibinfo{person}{Jens Dibbern}} (Eds.).
  \bibinfo{publisher}{Springer Berlin Heidelberg}, \bibinfo{address}{Berlin,
  Heidelberg}, \bibinfo{pages}{527--548}.
\newblock


\bibitem[\protect\citeauthoryear{Yao, Weden, Emerlyn, Zhu, and Kraut}{Yao
  et~al\mbox{.}}{2021}]%
        {Yao_Weden_Emerlyn_Zhu_Kraut_2021}
\bibfield{author}{\bibinfo{person}{Zheng Yao}, \bibinfo{person}{Silas Weden},
  \bibinfo{person}{Lea Emerlyn}, \bibinfo{person}{Haiyi Zhu}, {and}
  \bibinfo{person}{Robert~E Kraut}.} \bibinfo{year}{2021}\natexlab{}.
\newblock \showarticletitle{Together But Alone: Atomization and Peer Support
  among Gig Workers}.
\newblock \bibinfo{journal}{\emph{Proceedings of the ACM on Human-Computer
  Interaction}} \bibinfo{volume}{5}, \bibinfo{number}{CSCW2}
  (\bibinfo{year}{2021}), \bibinfo{pages}{1–29}.
\newblock
\urldef\tempurl%
\url{https://doi.org/10.1145/3479535}
\showDOI{\tempurl}


\bibitem[\protect\citeauthoryear{Zhang}{Zhang}{2007}]%
        {zhang2007toward}
\bibfield{author}{\bibinfo{person}{Ping Zhang}.}
  \bibinfo{year}{2007}\natexlab{}.
\newblock \showarticletitle{Toward a positive design theory: Principles for
  designing motivating information and communication technology}.
\newblock In \bibinfo{booktitle}{\emph{Designing information and organizations
  with a positive lens}}. \bibinfo{publisher}{Emerald Group Publishing
  Limited}.
\newblock


\end{thebibliography}

\end{document}